\documentclass[aps,prc,superscriptaddress,twocolumn]{revtex4-1}
\usepackage{epsfig}
\usepackage{graphicx}
\usepackage{bm}

\begin{document}
\title{Core-corona separation in the UrQMD hybrid model}


\author{J.~Steinheimer}
\affiliation{Institut f\"ur Theoretische Physik, Goethe-Universit\"at, Max-von-Laue-Str.~1,
D-60438 Frankfurt am Main, Germany}
\affiliation{Frankfurt Institute for Advanced Studies (FIAS), Ruth-Moufang-Str.~1, D-60438 Frankfurt am Main,
Germany}
\author{M.~Bleicher}
\affiliation{Institut f\"ur Theoretische Physik, Goethe-Universit\"at, Max-von-Laue-Str.~1,
D-60438 Frankfurt am Main, Germany}
\affiliation{Frankfurt Institute for Advanced Studies (FIAS), Ruth-Moufang-Str.~1, D-60438 Frankfurt am Main,
Germany}
\begin{abstract}
We employ the UrQMD transport + hydrodynamics hybrid model to estimate the effects of a separation of the hot equilibrated core and the dilute corona created in high energy heavy ion collisions. It is shown that the fraction of the system which can be regarded as an equilibrated fireball changes over a wide range of energies. This has an impact especially on strange particle abundancies. We show that such a core corona separation allows to improve the description of strange particle ratios and flow as a function of beam energy as well as strange particle yields as a function of centrality.
\end{abstract}

\maketitle

\section{Introduction}
The objective of the low energy heavy ion collider programs, at the RHIC facility on Long Island and the planned projects NICA in Dubna and FAIR near the GSI facility, is to find evidence for the onset of a deconfined phase \cite{Gyulassy:2004zy,Hohne:2005qm}. At the highest RHIC energies, experiments \cite{Adams:2005dq,Back:2004je,Arsene:2004fa,Adcox:2004mh} have already confirmed a collective behavior of the created (partonic) system, signaling a change in the fundamental degrees of freedom. Lattice QCD calculations indeed expect a deconfinement crossover to occur in systems created at the RHIC. As theoretical predictions on the thermodynamics of finite density QCD are quite ambiguous \cite{Fodor:2002km,Allton:2002zi,Allton:2003vx,Allton:2005gk,deForcrand:2003hx,Laermann:2003cv,DElia:2004at,DElia:2002gd}, one hopes to experimentally confirm a possible first order phase transition and consequently the existence of a critical endpoint, by mapping out the phase diagram of QCD in small steps.\\
Hadronic bulk observables which are usually connected to the onset of deconfinement are the particle flow and its anisotropies as well as particle yields and ratios \cite{Ollitrault:1992bk,Rischke:1996nq,Sorge:1996pc,Heiselberg:1998es,Scherer:1999qq,Soff:1999yg,Brachmann:1999xt,Csernai:1999nf,Zhang:1999rs,Kolb:2000sd,Bleicher:2000sx,Stoecker:2004qu,Zhu:2005qa,Petersen:2006vm,Gazdzicki:2004ef,Gazdzicki:1998vd}. It has often been proposed, that e.g. the equilibration of strangeness would be an indication for the onset of a deconfined phase, although this idea is still under heavy debate \cite{Koch:1986ud,Bratkovskaya:2000eu,BraunMunzinger:2003zz,Greiner:2004vm}.\\
The interpretation of experimental results and their relation to the deconfinement phase transition is most often circumstantial and extensive model studies are required to understand the multitude of observables. Therefore dynamical models for the description of relativistic heavy ion collisions are needed as input for the interpretation of the observed phenomena.\\
Dynamical approaches to heavy ion collisions are often based on two complementary theoretical concepts: the first being relativistic fluiddynamics \cite{Hofmann:1976dy,Stoecker:1986ci,Hung:1994eq,Rischke:1995ir,Aguiar:2000hw,Kolb:2003dz,Baier:2006gy,Song:2007ux,Schenke:2010nt,Werner:2010aa}. In this approach one assumes that the produced system can be described as an expanding liquid which is in local thermal equilibrium. The assumption of local equilibration is usually disputed which led to development of hydrodynamic models which employ viscous corrections. Apart from these complications, a general advantage of the hydrodynamic approach is that an equation of state, which contains the information on the active degrees of freedom of the system (potentially including a phase transition), can be easily introduced in the model.\\
The second type of models is based on the relativistic Boltzmann transport equation \cite{Molnar:2004yh,Xu:2004mz,Burau:2004ev,Bass:1998ca,Bleicher:1999xi,Cassing:2008sv,Linnyk:2010cr}. The applicability of this equation is independent of any equilibrium assumption which makes it superior to the hydrodynamic approach in this respect. However the implementation of a phase transition in such a microscopic model is far more complicated and still poses a great challenge to theorists.\\
To obtain a more comprehensive picture of the whole dynamics of heavy ion reactions various so called micro+macro hybrid approaches have been developed
during the last years \cite{Magas:2001mr} to combine the benefits of the two mutually complementary approaches discussed before. Here, one employs initial conditions that are calculated from a non equilibrium model followed by an ideal or viscous hydrodynamic evolution coupled to the Boltzmann equation for the final state \cite{Paiva:1996nv,Aguiar:2001ac,Teaney:2001av,Socolowski:2004hw,Hirano:2005xf,Hirano:2007ei,Bass:1999tu,Nonaka:2006yn,Dumitru:1999sf,Werner:2010aa,Song:2011qa}.\\

In this paper we present a study of the effects of the assumption of local thermal and chemical equilibrium in the description of heavy ion collisions at different beam energies. In particular we want to discuss cases where only parts of a created fireball can be regarded as being equilibrated. Following previous explorations such a system can be divided in a hot and dense core and a dilute corona \cite{Werner:2007bf,Becattini:2008yn}, where each part of the system should be treated on different theoretical footing. We will apply the UrQMD transport/hydrodynamics hybrid model as described in \cite{Petersen:2008kb} and modify it to allow for a consistent simultaneous description of a core-corona separated system.
Let us remark that the present model allows only to study the effects of local thermal and chemical equilibration, while it does not allow to pin down the actual dynamical processes which induce the early equilibration. In fact, the processes responsible for fast equilibration of the produced matter (probably instabilities or multi particle interactions \cite{Rebhan:2004ur,Greiner:2004vm}) still pose a great theoretical challenge in modeling heavy ion collisions.\\

In the following we will first explain in short the concept and implementation of the UrQMD hybrid model and how it is extend to allow for a consistent separation of the dense core and dilute corona of the fireball that is obtained. Then we present results that are sensitive to such a separation und discuss them in order before making concluding remarks.

\begin{figure}[t]
 \centering
\includegraphics[width=0.5\textwidth]{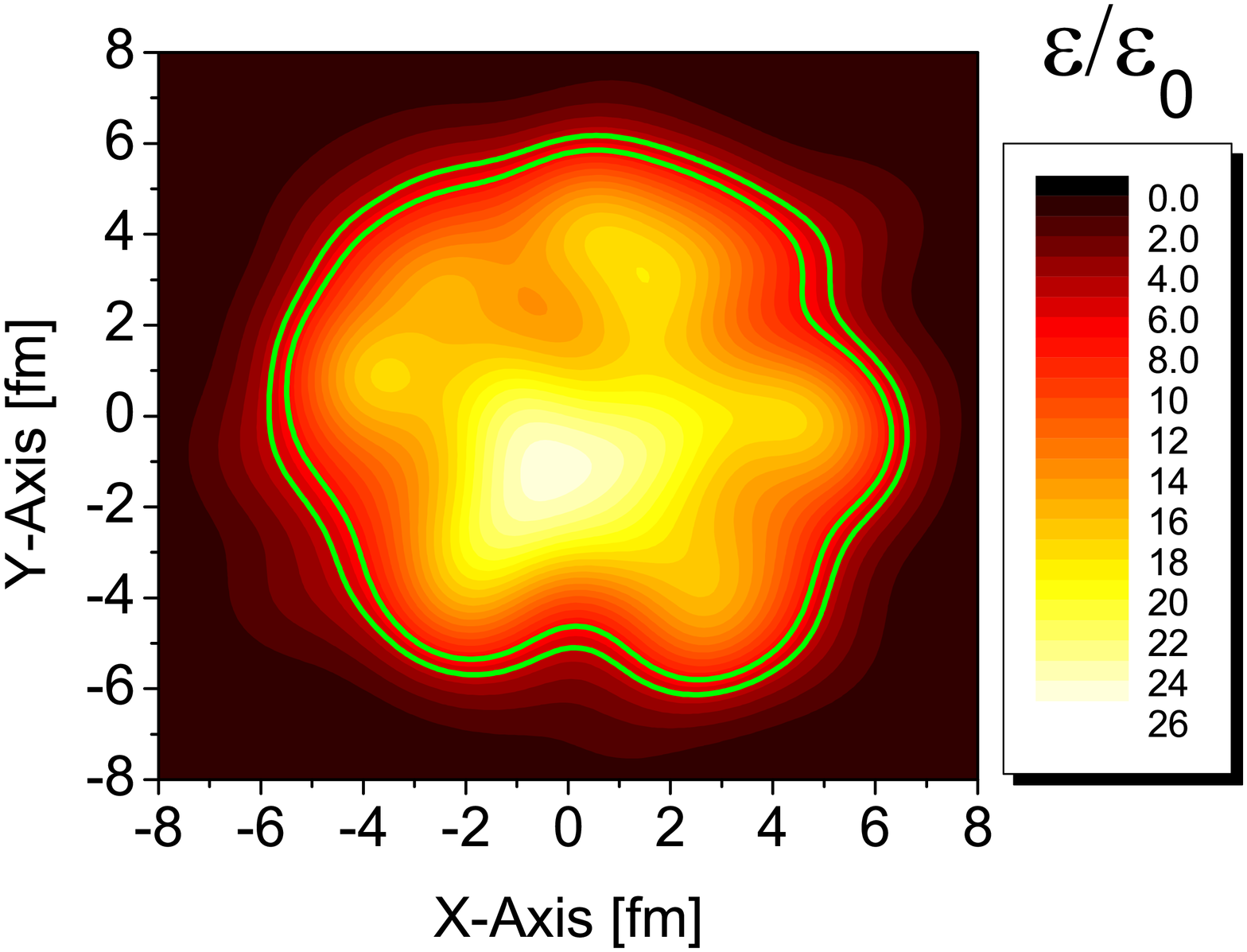}
 \caption{(Color online) Contour plot of the local rest frame energy density in the transverse plane ($z=0$) of a central ($b=0$) collision of Pb+Pb at $E_{lab}= 40 A$ GeV. The energy density is normalized to the ground state energy density ($\epsilon_0 \approx 145 \ \rm MeV/fm^3$). The two green lines correspond to lines of a constant energy density of $\epsilon/\epsilon_0 = 5$ and $7$.}
 \label{1}
\end{figure}

\section{The hybrid model}
Hybrid approaches to unite hydrodynamics and transport equations where proposed 10 years ago \cite{Bass:1999tu,Dumitru:1999sf} and have since then been employed for a wide variety of investigations \cite{Teaney:2001av,Socolowski:2004hw,Nonaka:2005aj,Hirano:2005wx}. The hybrid approach presented here is based on the integration of a hydrodynamic evolution into the UrQMD transport model \cite{Petersen:2008dd,Petersen:2008kb,Steinheimer:2007iy}. During the first phase of the evolution the particles are described by UrQMD as a string/hadronic cascade. Once the two colliding nuclei have passed through each other the hydrodynamic evolution starts at the time $t_{start}=2R/\sqrt{\gamma_{c.m.}^2-1}$, where $\gamma_{c.m.}$ denotes the Lorentz gamma of the colliding nuclei in their center of mass frame. While the spectators continue to propagate in the cascade, all other particles, i.e. their baryon charge densities and energy-momentum densities, are mapped to the hydrodynamic grid. By doing so one explicitly forces the system into a local thermal equilibrium for each cell. In the hydrodynamic part we solve the conservation equations for energy and momentum as well as the net baryon number current, while for the net strange number we assume it to be conserved and equal to zero locally. Solving only 
the equations for the net baryon number is commonly accepted in hydrodynamical models, although we have shown in earlier \cite{Steinheimer:2008hr}
publications that net strangeness may fluctuate locally. It is planned to also implement an explicit propagation for the net strange density.
Such an extension of the model also requires that the equation of state is extended in the net strange sector which is an investigation, currently 
underway and will be addressed in future publications.\\
The hydrodynamic evolution is performed using the SHASTA algorithm \cite{Rischke:1995ir} with an equation of state that incorporates a chiral as well as an deconfinement crossover and which is in agreement with thermodynamic results from lattice calculations \cite{Steinheimer:2010ib}. At the end of the hydrodynamic phase the fields are mapped to particle degrees of freedom using the Cooper-Frye equation \cite{Cooper:1974mv}. The transition from the hydrodynamic prescription to the transport simulation is done gradually in transverse slices of thickness 0.2 fm, once all cells in a given slice have an energy density lower than five times the ground state energy density (see also \cite{Steinheimer:2009nn}). The temperature at $\mu_B=0$ which corresponds to such a switching density
is roughly $T=170$ MeV which is close to what is expected to be the critical temperature. Detailed information of the transition curve in the phase diagram can be found in \cite{Petersen:2008dd}.
After this the final state interactions and decays of the particles are calculated and the system freezes out dynamically within the UrQMD framework.\\
For an extensive description of the model the reader is referred to \cite{Petersen:2008dd,Steinheimer:2009nn}.

\begin{figure}[t]
 \centering
\includegraphics[width=0.5\textwidth]{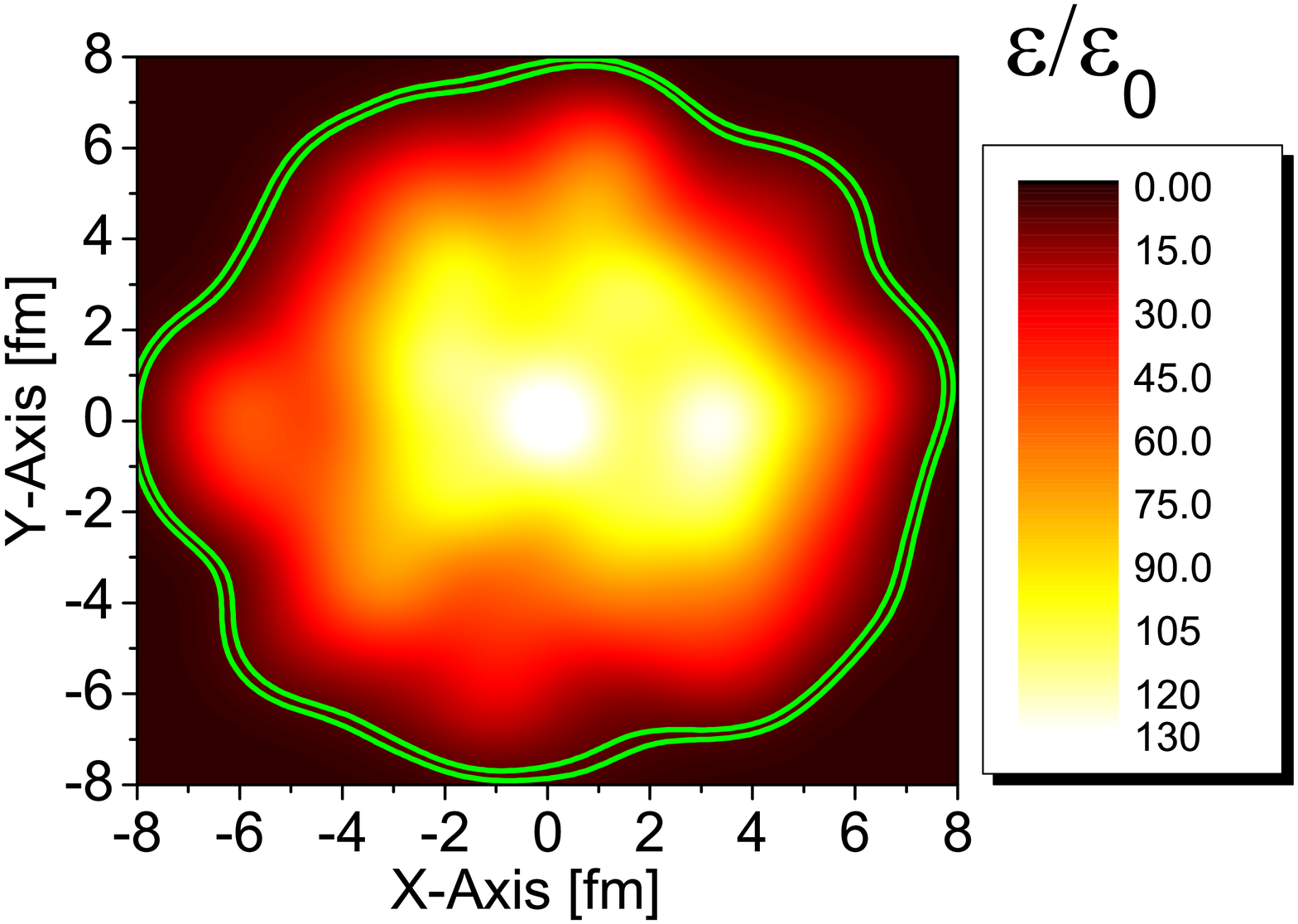}
 \caption{(Color online) Contour plot of the local rest frame energy density in the transverse plane ($z=0$) of a central ($b=0$) collision of Pb+Pb at $E_{lab}= 160 A$ GeV. The energy density is normalized to the ground state energy density ($\epsilon_0 \approx 145 \ \rm MeV/fm^3$). The two green lines correspond to lines of a constant energy density of $\epsilon/\epsilon_0 = 5$ and $7$.}
 \label{2}
\end{figure}

\section{Separating core and corona}
Essentially all previous hybrid model calculations have assumed that the whole system created enters a phase of local thermal equilibrium. As the local rest frame energy density varies in coordinate space, one would expect that some portions of the created system, already in the beginning of the evolution, have densities that are smaller than the transition energy density. In general, doing hydrodynamical simulations one neglects such portions of the system which never really enter an equilibrated phase. As a first step we want to estimate quantitatively how good such an assumption is. As a visualization of the problem, figures \ref{1} and \ref{2} show contour plots of the local rest frame energy densities (normalized to the ground state energy density) in the transverse plane ($z=0$) of central Pb+Pb collisions at two different energies. Figure \ref{1} depicts the distribution for $E_{lab}= 40 A$ GeV and figure \ref{2} for $E_{lab}= 160 A$ GeV. One can clearly see that the energy densities reached in the center of the system exceed the transition criterion for both cases , while the energy density for $E_{lab}= 160 A$ GeV is roughly 5 times that for $E_{lab}= 40 A$ GeV. The green solid lines are lines of constant energy density, $\epsilon = $ 5 and 7 times $\epsilon_0$. For the higher beam energy, almost all the system seems to have an energy density larger than this criterion while for $E_{lab}= 40 A$ GeV this is not the case, one observes that  parts of the system lie already outside of the hot and dense region.\\

In the following we describe how one can separate the system which is produced in the heavy ion collision in a dense core part, which will be propagated using the hydrodynamic prescription, and a dilute corona part for which we assume the UrQMD transport approach to be the correct model.\\
The idea that the fireball, created in a heavy ion collision, can be divided in a dense core which expands collectively and a dilute corona which is dominated by hadronic scatterings is not new. The first time this idea was adapted in a dynamical model for heavy ion collisions was in \cite{Werner:2007bf}. In this approach the system was divided in transverse cells of a certain pseudorapidity range. Whenever the transverse string density in such a cell was above a certain value then it was considered a part of the equilibrated core otherwise was is considered part of the corona. In a different approach the core-corona separation was made, using a Monte Carlo Glauber model and distinguishing between nucleons that have interacted once or more than once, while those nucleons which have interacted only once were regarded as Corona part \cite{Aichelin:2008mi,Becattini:2008ya,Aichelin:2010ns,Aichelin:2010ed}.\\
For the present study we will apply a method similar to that used in \cite{Werner:2007bf}. At the time $t_{start}$ of the transition from the UrQMD model to the hydrodynamic phase we calculate the scalar constituent quark number density (Mesons count 2 for $q+\overline{q}$ and Baryons count 3 for $q+q+q$) at the position of every particle. This is done by describing every hadron as a Lorentz contracted Gaussian distribution of its constituent quark number and then sum over the contributions of all particles to a given space point. As a result one obtains the quark number density $\rho_q$ at the position of every hadron in the UrQMD model. If the density at the particles position is above a certain critical separation density it is used to calculate the initial density profiles for the hydrodynamic evolution as outlined above. If the density is below the separating density it will remain in the transport model and will be propagated there in parallel to the hydrodynamic evolution. After the transition from the hydrodynamic phase, back to the transport model occurs, all particles can then again re-interact and decouple dynamically.res, all particles can then again re-interact and decouple dynamically.
Note that this procedure is similar to tho one applied in \cite{Werner:2007bf}, although a distinct difference is that we calculate the scalar density at every particles position in coordinate space. As for higher energies, the particle density should be roughly independent of the pseudorapidity (boost invariance), the definition of a corona via the transverse string density as in \cite{Werner:2007bf} is sufficient. For lower energies this relation does not hold anymore and the full calculation of the particle number density seems more appropriate. On the other hand, the present approach becomes unfeasible at some point as particles from all rapidities contribute to the local density of any other particle. One might therefore, optionally, apply a cut in pseudo-rapidity (of $\Delta \eta = 0.5$) for particles which contribute to the local density $\rho_q (\vec{x})$.\\
Our procedure introduces the density $\rho_q$ as another parameter in the model. In the present investigation we will constrain this parameter to lie between 4 and 5 times $\rho_{q0}$ (where $\rho_{q0}= 0.15 \cdot 3 \ fm^{-3}$, the ground state quark density at $T=0$). This choice of parameter is taken, because we try to keep the cut off density to enter the hydrodynamic phase close to the density criterion for the transition from the hydrodynamic phase back to the hadronic afterburner. The values of 4 and 5 times $\rho_{q0}$ closely correspond to the energy densities of 5 and 7 times $\epsilon_0$, when we consider a hadron resonance gas, which includes the same degrees of freedom as does UrQMD.\\

\section{Energy dependence}

\begin{figure}[t]
 \centering
\includegraphics[width=0.5\textwidth]{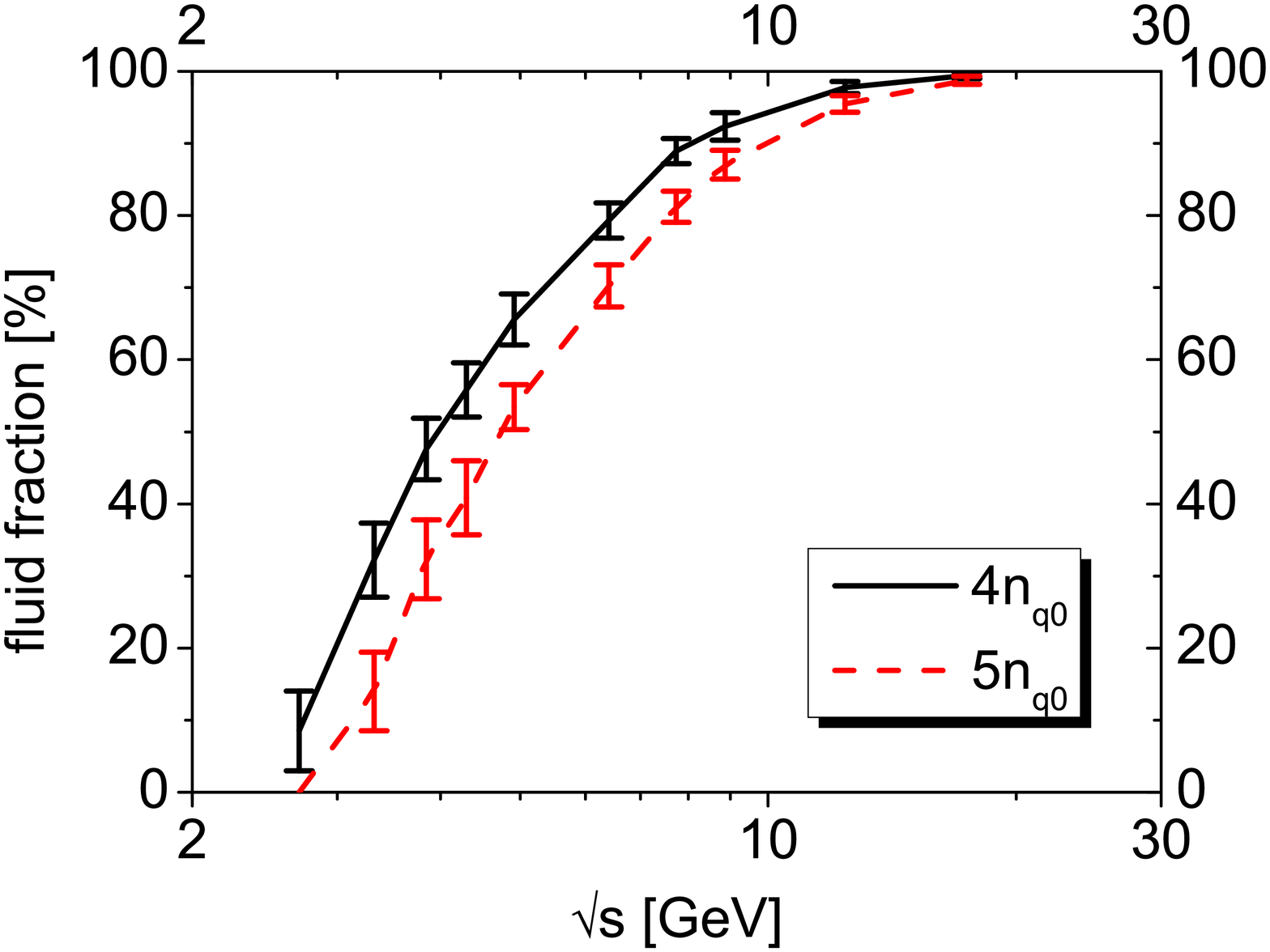}
 \caption{(Color online) Fraction of the total energy of the system which is transferred into the hydrodynamic phase as a function of center of mass beam energy, for central ($b<3.4$ fm) collisions of AuAu/PbPb . The black line corresponds to a cut of energy density of $\rho_q=4 \rho_{q0}$ and the red dashed line to $\rho_q=5 \rho_{q0}$. The error bars indicate the mean deviation of the fluid fraction on an event by event basis.}
 \label{3}
\end{figure}

In this section we will concentrate in the beam energy dependence of effects of a core-corona separation  as described above. We will apply the model for most central ($b<3.4$ fm) collisions of AuAu/PbPb at different beam energies and present results for choices of the density cut off parameter of 4 and 5 time $\rho_{q0}$.\\
Figure \ref{3} shows the fraction of the total energy of the colliding system (excluding spectators) which is transferred into the hydrodynamic phase. For the lowest beam energy, $E_{lab}=2 A$ GeV, only a vanishing fraction of the system can be regarded as being in local thermal equilibrium. The fraction increases slowly with the beam energy, while only at the highest SPS energies one can regard the whole system as being equilibrated. Changing the density cut off parameter $\rho_q$ only results in a small shift at intermediate beam energies, while at the highest SPS energies the density gradients of the produced system are so large that the result is insensitive on the exact value of the $\rho_q$ parameter. The 'error' bars in the figure represent the mean deviation of the fluid fraction on an event-by-event basis.\\

\begin{figure}[t]
 \centering
\includegraphics[width=0.5\textwidth]{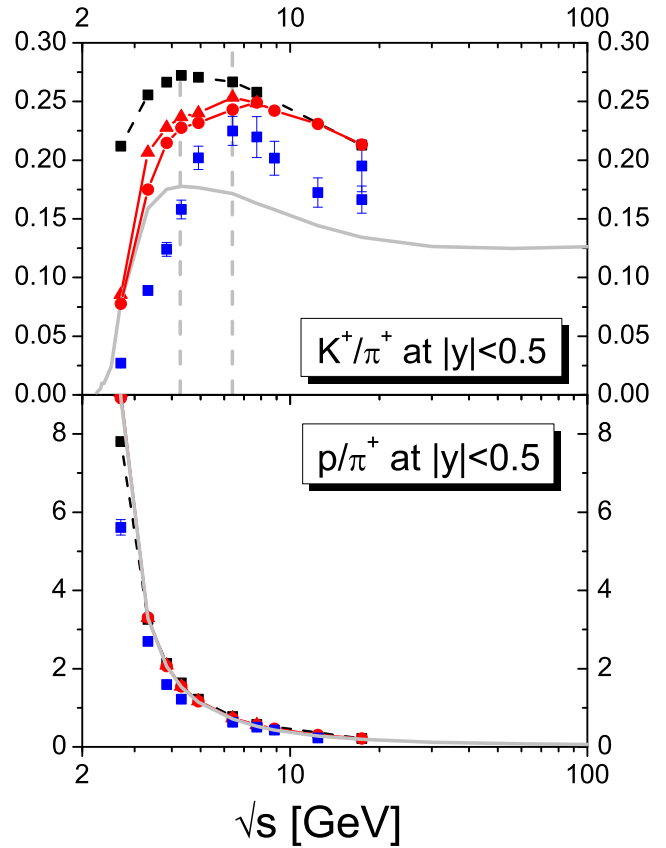}
 \caption{(Color online) Particle ratios of protons to pions, upper plot, and positively charged kaons to pions, lower plot. The results are for the mid rapidity region ($|y|<0.5$) of central ($b<3.4$ fm) collisions of AuAu/PbPb ions. The different model results (lines are explained in the text) are compared to experimental data. The horizontal dashed lines are shown to point out the peak positions of the $K^+/\pi^+$ ratio for the standard UrQMD calculation and the data.}
 \label{4}
\end{figure}

\begin{figure}[t]
 \centering
\includegraphics[width=0.5\textwidth]{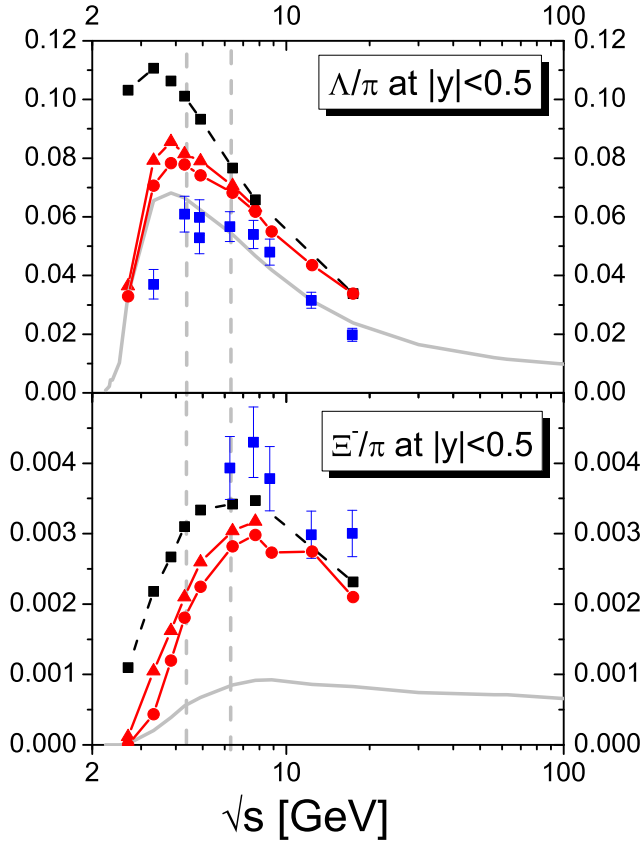}
 \caption{(Color online) Particle ratios of lambdas to pions, upper plot, and negatively charged Xi's to pions, lower plot. The results are for the mid rapidity region ($|y|<0.5$) of central ($b<3.4$ fm) collisions of AuAu/PbPb ions. The different model results (lines are explained in the text) are compared to experimental data. The horizontal dashed lines are shown to point out the peak positions of the $K^+/\pi^+$ ratio for the standard UrQMD calculation and the data.}
 \label{5}
\end{figure}

As a next step we investigate the dependence of experimental hadronic observables like particle yields and flow, on the separation procedure. Figure \ref{4} depicts the beam energy dependence of the ratios of protons to pions (upper plot) and positively charged kaons and pions (lower plot) in the mid-rapidity region of central $b<3.4$ heavy ion collisions ($|y|<0.5$). Here, as well as in the following plots, for comparison the solid grey line represents the results of the default UrQMD calculation without an intermediate hydrodynamic stage. The black dashed line with square symbols depicts the hybrid model results without any core-corona separation, which means that the whole fireball is regarded as being in local thermodynamic equilibrium. The red solid lines show the results obtained when we apply our core-corona separation, using the two different values of the density cut off parameter (triangles: $\rho_q = 4 \cdot \rho_{q0}$, circles: $\rho_q = 5 \cdot \rho_{q0}$). The model results are compared to data from different experiments \cite{Alt:2008qm,Afanasiev:2002mx,Ahle:1999uy,Adams:2003xp,Adcox:2003nr,Klay:2003zf,:2007fe,Mitrovski:2006js} which are shown as blue square symbols.\\
For the non-strange protons and pions the assumption of local thermal equilibration seems not to change the results on the particle ratio and all different model calculation give a rather good description of the data.\\

\begin{figure}[t]
 \centering
\includegraphics[width=0.5\textwidth]{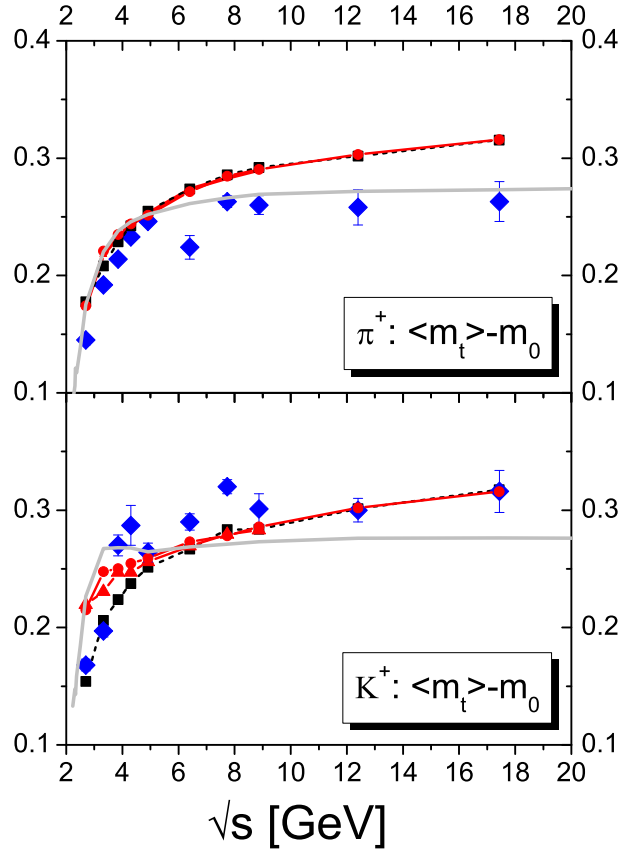}
 \caption{(Color online) Excitation functions of the mean transverse mass, for the mid rapidity region ($|y|<0.5$) of central ($b<3.4$ fm) collisions of AuAu/PbPb ions. Positively charged pions are shown in the upper panel and kaons in the lower panel. Experimental data is shown as diamond symbols \cite{Ahle:1999uy,:2007fe,Afanasiev:2002mx} (lines are explained in the text).}
 \label{7}
\end{figure}

However, if one looks at the ratio of the positively charged kaons to pions we observe considerable differences in the results obtained from the different approaches. Generally, the assumption of thermal equilibrium drastically enhances the production of strange particles when compared to the UrQMD non-equilibrium approach. Especially for the very low beam energies this leads to a drastic overestimation of the ratios involving strange particles in the standard UrQMD hybrid model.  
In contrast using the core-corona separation approach the fraction of the system, for which local equilibrium is assumed, changes with beam energy. Therefore, for the lowest energies, one smoothly recovers the default UrQMD results. At intermediate energies the core-corona result is generally in between the transport model and default hybrid model. It should be noted that the position of the peak in the $K^+/\pi^+$ ratio depends on the core fraction and only coincides for the new core-corona approach with the available data. For the default hybrid and transport model calculations, the peak appears at lower energies. As a side remark let us state that in the usual transport simulations the position of the peak is solely determined by the transition from a baryon to meson dominated system, while for the present core-corona approach the slow onset of strangeness equilibration plays the driving role.\\
 
The beam energy dependence of strange baryon to pion ratios is shown in figure \ref{5}, where the vertical dashed lines again indicate the positions of the peaks in the $K^+/\pi^+$ ratio. The description of the $\Lambda/\pi$ is again improved, when one assumes that only a part of the system is fully equilibrated.
Fore the $\Xi^-/\pi$ ratio (lower plot in figure \ref{5}) the default UrQMD calculation drastically underpredicts the production rate of the multistrange baryon. Even the default hybrid model result seems to underestimate the data slightly and this ratio even decreases in our new core-corona approach. However, the effect is smaller than for the single strange hadrons.
For both strange baryon ratios the peak position is found to be independent of the applied model parametrization.\\

Next we turn to the investigation of average particle flow. Even more than particle yields, their momenta and collective motion could depend on the degree of equilibration in the system. Figure \ref{7} shows the excitation functions of the mean transverse masses of pions and kaons compared to data \cite{Ahle:1999uy,:2007fe,Afanasiev:2002mx}. For the positively charged pions we observe almost no dependence on the parametrization of the core-corona separation, in the hybrid model.
The $K^+$ excitation function (shown in the lower part of figure \ref{7}) shows small differences at low beam energies. Here the mean transverse mass is increased in the non-equilibrium transport approach as compared to the hybrid model calculation. Interestingly, the description of the $\pi^+$ data at high energies is better in the default UrQMD approach, while for the $K^+$ it is better in the hybrid model. 

In \cite{Petersen:2009mz} it has been pointed out, that the surplus of low momentum pions in the data as compared to the hybrid model calculations, which would lead to a decrease in the $K^+/\pi^+$ and the mean $m_T$ of pions, can be contributed to heavy resonance contributions as well as non-equilibrium corrections to the hydrodynamic phase.\\

\begin{figure}[t]
 \centering
\includegraphics[width=0.5\textwidth]{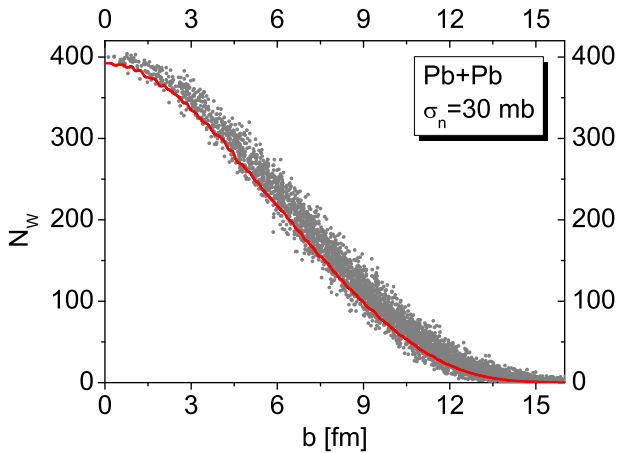}
 \caption{(Color online) The dependence of the number of wounded nucleons $N_W$ as a function of the impact parameter for collisions of heavy ions at $E_{lab}= 40 A$ GeV. The red line shows the result from a Monte Carlo Glauber simulation, assuming an inelastic nucleon-nucleon cross section of $\sigma_n=30$ mb. The points depict the results from the hybrid model calculation on an event-by-event basis.}
 \label{8}
\end{figure}

\section{Centrality dependence}

Instead of varying the temperature and density by a beam energy scan, such a variation could also be achieved by changing the centrality of the collision. In our calculation this can be done by changing the impact parameter b. In experiment, as the determination of the impact parameter is usually not straight forward, one usually gives observables as function of the number of wounded nucleons. This is the number of nucleons which have undergone a primary binary collision and their energy can therefore contribute to the fireballs total energy.\\
In the transport and hybrid model calculations, the definition of the number of wounded nucleons is by no means trivial. Late time secondary interactions, which would only excite the spectator fragment, can remove spectator nucleons from the fragment leading to an overestimation of $N_W$. We therefore define the number of wounded nucleons in our model calculations, as the number of nucleons which have not interacted until the time $t_{start}$ (see definition above). In this way one obtains a dependence of $N_W$ on the impact parameter $b$ which is in agreement with the estimate of a Monte-Carlo Glauber model calculation \cite{Bialas:1976ed,Eskola:1988yh} (see figure \ref{8}). Such a Glauber model is often used to estimate $N_W$ from experimental data. Therefore we can compare our results with experiment, using our definition of $N_W$ without invoking the real experimental trigger conditions.\\

\begin{figure}[t]
 \centering
\includegraphics[width=0.5\textwidth]{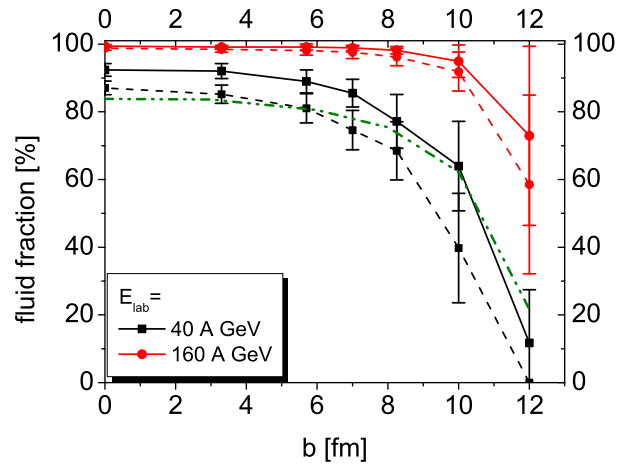}
 \caption{(Color online) Fraction of the total energy of the system which is transferred into the hydrodynamic phase as a function of the impact parameter b, for collisions of Pb nuclei. The red lines with circles depict the results for $E_{lab}= 40 A$ GeV and the black lines with squares for $E_{lab}= 160 A$ GeV. We compare results for both choices of the core density parameter $\rho_q= 4$ (solid lines) and $5 \rho_{q0}$ (dashed lines). If a cut in pseudo rapidity as applied for the calculations at  $E_{lab}= 160 A$ GeV we obtain the green dash dotted line as a result for $\rho_q=5 \rho_{q0}$. The error bars indicate the mean deviation of the fluid fraction on an event by event basis.}
 \label{9}
\end{figure}

Let us start with an investigation of the fluid fraction as a function of the impact parameter. Figure \ref{9} displays the fluid fraction for two different energies as a function of the impact parameter $b$. For each energy the results for the two different cut off densities, where the solid lines correspond to $\rho_q=5 \rho_{q0}$ and the dashed lines to $\rho_q=4 \rho_{q0}$. The green dash dotted line shows the effect of the cut in pseudorapidty in the definition of the density for $E_{lab}=160 A$ GeV.\\
For both energies one observes a dependence of the fluid fraction on the input parameter. While at $E_{lab}= 40 A$ GeV (red lines with circles) this dependence is rather strong, it is rather weak for $E_{lab}= 160 A$ GeV (black lines with squares). If a cut in $\eta$ for the calculation of the local density is applied, the impact parameter dependence becomes much stronger for th highest SPS energy (green dash-dotted line). 

\begin{figure}[t]
 \centering
\includegraphics[width=0.5\textwidth]{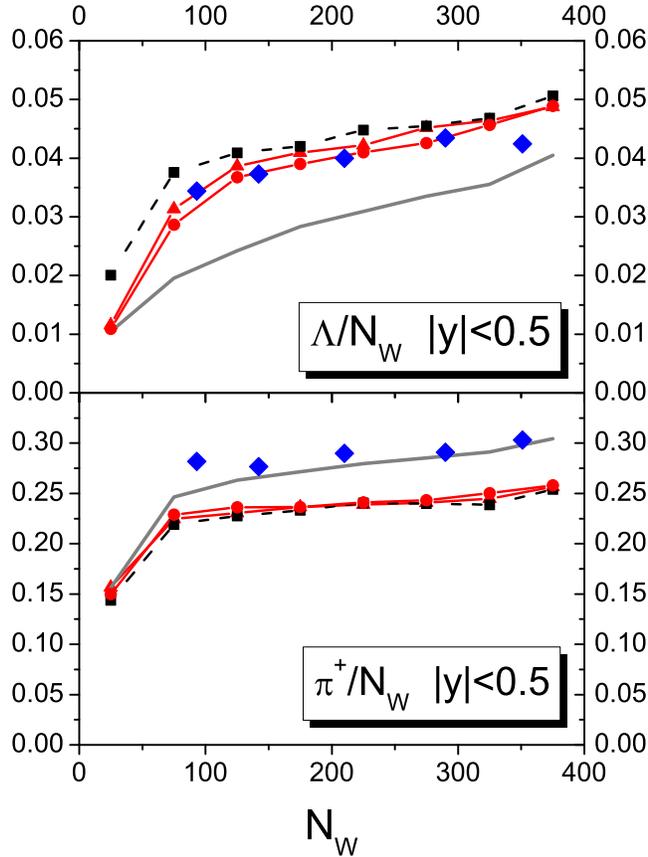}
 \caption{(Color online) Centrality dependence of the number of $\Lambda$'s (upper plot) and pions (lower plot) per wounded nucleon, produced in heavy ion collisions at $E_{lab}=40 A$ GeV, compared to data. Shown is the yield at mid rapidity $|y|<0.5$. 
  The grey line is the result from the default UrQMD model and the dashed black line with squares depicts the default hybrid model results. The solid red lines show the calculations with the core-corona separated hybrid model with two different values of the core density parameter $\rho_q= 4$ (triangles) and $5 \rho_{q0}$ (circles).}
 \label{10}
\end{figure}

\begin{figure}[t]
 \centering
\includegraphics[width=0.5\textwidth]{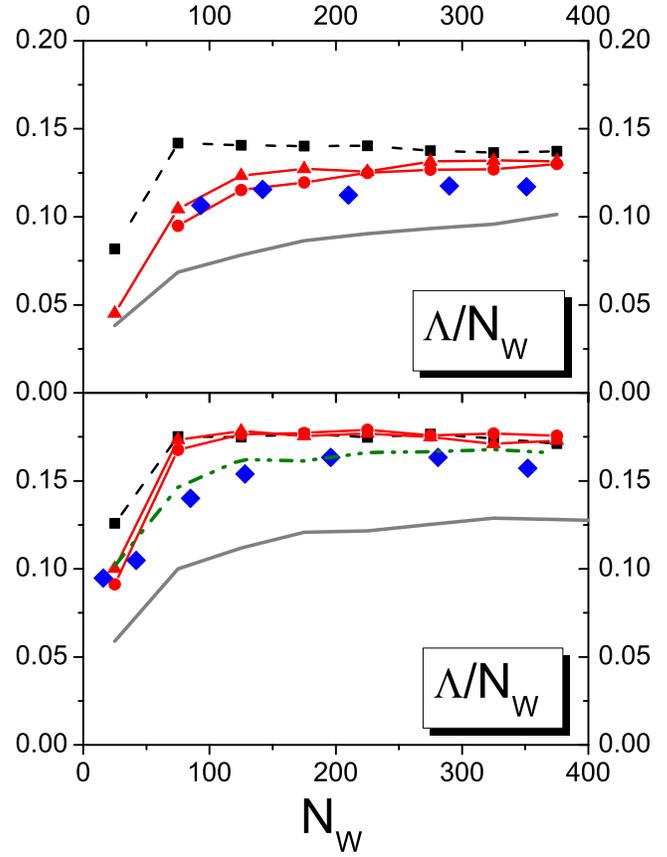}
 \caption{(Color online) Centrality dependence of the number of $\Lambda$'s per wounded nucleon produced in heavy ion collisions at $E_{lab}=40 A$ GeV (upper plot) and $E_{lab}=160 A$ GeV (lower plot), compared to data. Shown is the $4 \pi$ integrated yield. The line styles are as in the previous figure. If a cut in pseudo rapidity as applied for the calculations at  $E_{lab}= 160 A$ GeV we obtain the green dash dotted line as a result for $\rho_q=5 \rho_{q0}$.}
 \label{11}
\end{figure}

In the last part of this paper we will discuss the results on the centrality dependence of different hadronic observables and compare them to data \cite{Anticic:2009ie,Alt:2008qm,Alt:2004wc,Alt:2004kq,Alt:2003ab} (the centrality selection with regard to the number on wounded nucleons is taken from \cite{Petersen:2009zi}). Figure \ref{10} displays the mid-rapidity ($|y|<0.5$) yields of pions and lambdas divided by the number of wounded nucleons, as a function of $N_W$. Here we the analysis is restricted to results for $E_{lab}=40 A$ GeV due to stronger dependence of the fluid fraction on the centrality at this energy. The different lines depict the results for the default UrQMD model in its cascade mode (grey line), the default hybrid model (black dashed line with square symbols) and the different parametrizations of the core-corona separated hybrid model (red lines with symbols). For the most central collisions all models reproduce the data equally well, while the centrality dependence can only be captured by the hybrid model, and especially the core-corona separated versions.\\
The pion yield on the other hand shows only a weak sensitivity on the presence of a corona and is well reproduced with the UrQMD cascade version.\\

While figure \ref{10} only displays the results at $E_{lab}=40 A$ GeV, figure \ref{11} shows a comparison of the centrality dependence of the $\Lambda$ multiplicity for $E_{lab}= 40$ and $160 A$ GeV. For the lower energy the core-corona separated version again gives the best result. While at the highest SPS energy, all hybrid model results show almost no centrality dependence, which is in contrast to experimental data. it seems that our definition of the corona begins to fail when the beam energy becomes so large that the produced system can be regarded as being boost invariant. To obtain a rapidity independent density, we apply an additional cut in $\eta$, as described above, for the calculation of $\rho_q$. This way one effectively reduces the system to 2 dimensions and one obtains a result which is comparable to that of the mere geometrical picture proposed in \cite{Werner:2007bf,Aichelin:2008mi,Becattini:2008ya,Aichelin:2010ns,Aichelin:2010ed}.

\begin{figure}[t]
 \centering
\includegraphics[width=0.5\textwidth]{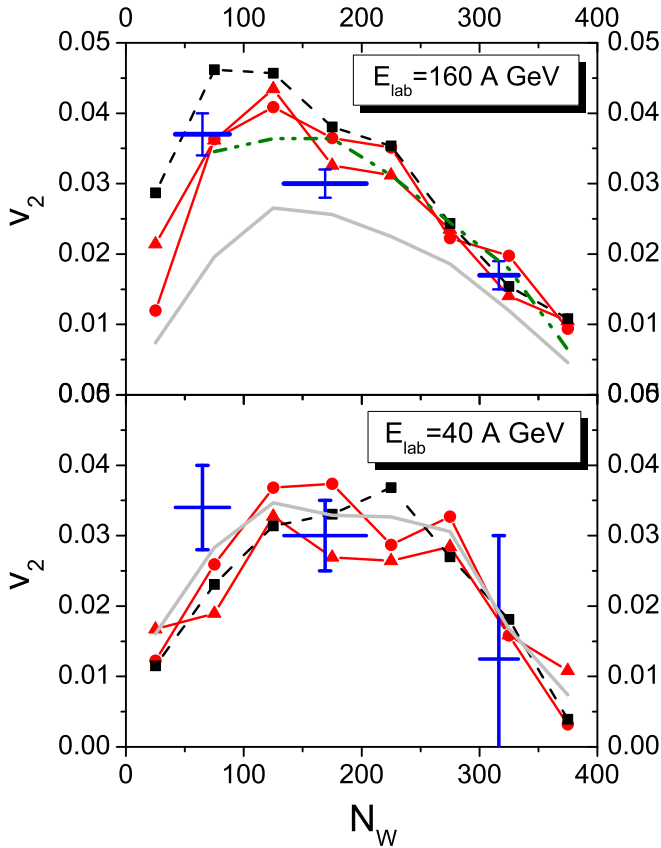}
 \caption{(Color online) Centrality dependence of the charged particle elliptic flow at mid rapidity $|y|<0.5$ for collisions of Pb at beam energies of $E_{lab}= 40$ (lower plot) and $160 A$ GeV (upper plot). The line styles are as in the previous figure.}
 \label{12}
\end{figure}

Figure \ref{12} summarizes our results on the centrality dependence of the elliptic flow of charged pions, with $v_2=\left\langle \cos\left[2(\phi- \Phi_R)\right]\right\rangle $, where $\Phi_R$ denotes the reaction plane. Again both beam energies $E_{lab}= 40$ and $160 A$ GeV are depicted. A general challenge for the experimental determination of $v_2$ is the correct determination of the reaction plane which sets the coordinate system in which the elliptic flow is defined. The experimental systematic uncertainty is reflected in large error bars and is especially severe at the lower energies. Nevertheless our calculations show that the value of $v_2$ is hardly sensitive on the approach which is used for the calculations. All results, from the default UrQMD and the hybrid model, essentially give the same centrality dependence of $v_2$ at $40 A$ GeV.
At the highest SPS beam energy the picture is already different. Here the default UrQMD results underestimate the elliptic flow, while the default hybrid model overestimates it. Again, the core-corona separated version of the hybrid model improves the description.\\

\section{Conclusion}  
We presented a method to explore the dynamics of the system produced in high energy heavy ion collisions and to effectively divide it into an equilibrated core and a dilute corona. To this aim the UrQMD hybrid approach was applied, where the dense and equilibrated core is described hydrodynamically and the dilute corona by the non-equilibrium transport approach. To distinguish between the two separated regions we employed a local particle density criterion.\\
It was found that the fraction of the system which can be regarded as being in local thermal and chemical equilibrium slowly increases in the energy range between $E_{lab}= 5-40 A$ GeV. While observables of non-strange hadrons appeared to be insensitive on this separation, strange hadron properties showed considerable modifications. Strange particle yields as well as their radial flow and especially the description of the 'horn' in the $K^+/\pi^+$ ratio were improved. From this point of view we explained the drastic increase in the $K^+/\pi^+$ up to $E_{lab}= 20-30 A$ GeV with an onset of chemical equilibration of strangeness. In thermal models which also explain the steep increase in strangeness production \cite{Andronic:2005yp,Andronic:2008gu,Becattini:2003wp,Becattini:2000jw} canonical corrections or the introduction of a strangeness saturation parameter, are usually responsible for the suppression of strangeness at low energies.\\

If the rapid equilibration of strangeness is caused by a change in the properties of the active degrees of freedom, present in the initial state of the collision,then the present study suggests an onset of such a new phase in reactions at beam energies of $E_{lab}\approx 5-40 A$ GeV. In fact the change of the properties of the Matter involved would not change suddenly at some specific beam energy but rather continuously over a wide range of collision energies.\\

In the second part of this paper we discussed the centrality dependence of different hadronic observables within the core-corona separated approach. Again the description of strange hadron observables is improved in the core-corona version, when compared to the default hybrid model. On the other hand the hybrid model gives no good description the excitation function of the mean transverse mass of pions and centrality dependence of pion multiplicities, independent of the core-corona separation. 
For our model we apply an ideal fluid dynamical description without viscous corrections. This is a possible origin for both observations, as viscosity leads to entropy production which is directly related to the pion production rate. Furthermore it can account for a decrease in the average flow in the hydrodynamic phase. Another contribution to the too small pion yield, as well as the too large average momentum are missing contributions from high mass (mesonic) resonances, decaying predominantly into pions, which are not explicitly included in the Cooper-Frye transition and the transport model (as well as in the UrQMD model).

 At the highest SPS energy we only obtain a very moderate centrality dependence of the core fraction of the system. This is somewhat in contrast to studies where the core fraction is calculated only on a 2 dimensional projection of the system on the transverse plane, indicating that the system produced at $E_{lab} = 160 A$ GeV seems to have a rapidity independent density. As the Lorentz contraction of the incoming nuclei is also rather strong, a Lorentz invariant formulation of the density, as well as the hydrodynamical equations seems more suitable. To approximate a rapidity independent density, we applied a cut in $\eta$ for the definition of the local particle density leading to a considerable improvement of the centrality dependence of strange particle yields which supports a definition of the core as proposed in \cite{Werner:2007bf,Aichelin:2008mi}, at least at energies above $E_{lab} \approx 160 A$ GeV. However, at lower energies a full 3 dimensional evaluation of the created system is in order.\\

\section*{Acknowledgments}
This work was supported by BMBF, HGS-hire and the Hessian LOEWE initiative through the Helmholtz International center for FAIR (HIC for FAIR). The authors thank D. Rischke for providing the SHASTA code and Dr. H. Petersen for fruitful discussions. The computational resources were provided by the Frankfurt LOEWE Center for Scientific Computing (LOEWE-CSC).

\end{document}